\documentclass[twocolumn,showpacs,preprintnumbers,amsmath,amssymb]{revtex4}

\usepackage{graphicx}
\usepackage{dcolumn}
\usepackage{bm}

\newcommand\beq{\begin{eqnarray}}
\newcommand\eeq{\end{eqnarray}}
\newcommand\la{\langle}
\newcommand\ra{\rangle}

\input{epsfig.sty}

\begin{document}


\vspace*{0.5cm}


\title{Charmed-meson spectroscopy in QCD sum rule}

\author{A.~Hayashigaki$^{1}$ and K. Terasaki$^{2}$}
\affiliation{
$^{1}$ Department of Physics, Kyoto University, Kyoto 606-8502, Japan}
\email{arata@ruby.scphys.kyoto-u.ac.jp}
\affiliation{
$^{2}$ Yukawa Institute for Theoretical Physics, Kyoto University, 
Kyoto 606-8502, Japan}
\email{terasaki@yukawa.kyoto-u.ac.jp}

\date{\today}

\begin{abstract}
We elaborate masses of open-charm mesons ($c\bar{d}$ and $c\bar{s}$)
with $J^{P}=0^{-},1^{-},0^{+},1^{+}$ from a viewpoint of ordinary
light-heavy systems in the analysis of standard Borel-transformed QCD
sum rule.  This analysis is implemented with the operator product
expansion up to dimension 6, permitting corrections to the order
$\alpha_s$ and to the order $m_s$, and without relying on
$1/m_c$-expansion. 
The obtained results following our stringent criteria for the continuum-threshold dependence, the Borel window and the Borel stability,
indicate that the charmed-meson masses in the $0^{+}$ channel are
overestimated by $100 \sim 200$ MeV in comparison with the experimental data, 
which were recently reported as the rather broad state of
$D^{*+}(2351)$ 
and the extremely narrow state of $D_s^{*+}(2317)$, respectively. 
Such large mass-discrepancies from the data cannot be seen in other
channels, where conversely our results of the $c\bar{s}$-meson masses
are even underestimated somewhat in comparison with data, 
independent of the value of the strange-quark mass adopted 
in our calculations. 
From these results, it might be expected that the measured low mass 
of $D_s^{*+}(2317)$ is a manifestation of an exotic state 
with the structures of a four-quark or a molecule, 
while at present the $D^{*+}$ is not in conflict with the existing data
due to its large width.
\end{abstract}


\maketitle


\section{Introduction}
The detailed study of charmed-meson spectroscopy has provided us 
with understanding of not only its structure and property peculiar 
to the light-heavy system, but also widely the basic property 
of the strong interactions.
Especially, the charmed mesons have played a part for 
the success of a potential model analysis in meson spectroscopy 
through a unified framework \cite{GI}.
However, recent experimental discovery of scalar 
and axial-vector $c\bar{s}$-mesons
\cite{BABAR1,BABAR2,CLEO1,BELLE1,FOCUS1}, 
which have been only missing $p$-wave states \cite{BS}, 
cast a shadow on such a success of the potential models, 
because their predictions \cite{GK,DE,CJ} give 
heavier masses by $100\sim 200$ MeV than the observations. 
Similar results have been seen also in QCD sum rule coupled 
with heavy quark effective theory (HQET) \cite{DHLZ} 
and lattice QCD simulations \cite{Bali,UKQCD}.
In the meanwhile, an interpretation of states with $J^P=0^+, 1^+$
as the chiral partner of ground states with $0^-, 1^-$
in the heavy quark limit 
requires relatively lower masses for the ($0^+, 1^+$) states 
than the above predictions. The analysis based on such idea
gives the results comparable to the observations \cite{NRZ,BEH}.
This problem with the mass discrepancy has triggered 
a lot of debates on the natures of the $c\bar{s}$ mesons 
in the viewpoints of modification of theoretical models, 
{\it e.g.} involving $DK$-mixing \cite{vBR}, 
or introduction of unconventional structures 
like two-meson molecular states \cite{BCL,Szczepaniak} 
or four-quark states \cite{CH,Terasaki} or their mixture \cite{BPP}.
There is, however, still no definite explanation for the discrepancy. 

In this paper, we study eight states 
of the charmed mesons with $J^{P}=0^{-}, 1^{-}, 0^{+}, 1^{+}$,
which are composed of the corresponding four states of $c\bar{s}$-mesons 
and similarly four states of $c\bar{n}$-mesons with $n=u,d$.
The masses of their states are elaborated through 
all the same analysis from the viewpoint of the conventional 
interpretation as light-heavy-quark bound systems.
We then focus attention on the masses of two $c\bar{s}$-mesons with $0^+$ and
$1^+$ in comparison with the experimental data. 
This method would be quite useful to extract information 
on specific states like $0^+$ and $1^+$ 
in the analysis involving relatively large theoretical uncertainties.
Here, in the $1^+$ channel, we deal with only a channel of 
charge-conjugation parity $+1$, {\it i.e.} $J^{P(C)}=1^{+(+)}$ channel, 
although indeed the light-heavy system does not have this parity 
as a good quantum number and thus takes place a mixing 
between $1^{+(+)}$ and $1^{+(-)}$ channels.

In the present work, such a physical motivation can be realized 
technically by relying on the standard 
QCD sum rule (QSR) approach \cite{SVZ,Narison2}, 
which evaluates hadron properties by using the correlator 
of the quark currents over the physical vacuum. 
We perform its calculation without taking $1/m_c$-expansion, 
where $m_c$ is the charm quark mass. 
It is because the charm quark is comparatively light 
and some properties of charmed-meson systems seem to be not 
necessarily valid for such an expansion, especially for their masses.
However, some analyses have approximated at the first few 
terms of the expansion based on the HQET \cite{CJ,DE,DHLZ,Bali}.
Our analysis is implemented with the operator product expansion 
up to dimension 6, permitting corrections to the order $\alpha_s$ 
and to the order $m_s$, where $\alpha_s$ is strong coupling constant 
and $m_s$ the strange-quark mass.
In Borel-transformed QSR, so-called Borel sum rule (BSR) \cite{SVZ}, 
we extract charmed-meson masses with 
imposing criteria for the continuum-threshold dependence, 
the Borel window and the Borel stability. 
Indeed, we find that the $D(0^+), D_s(0^+)$-meson masses are
overestimated by $100 \sim 200$ MeV compared with the experimental data, 
while such a tendency cannot be seen in other channels.
Conversely, in other channels the $c\bar{s}$-meson masses are even
underestimated somewhat in comparison with the data, 
independent of the value of the strange-quark mass adopted in our analysis.
Accounting for the measured broad width of the $D(0^+)$-meson, 
the state seems to be not in conflict with the existing data. 
From these results, it is quite possible that the measured 
low mass of $D_s^{*+}(2317)$ is a manifestation of any exotic states 
with the structures of a four-quark or a molecule.

This paper is organized as follows.
In sec.~II, we first mention the current status of charmed-meson 
spectroscopy in more detail, from both experimental and theoretical
aspects. In sec.~III, we formulate the general form of BSR 
and derive the analytic forms of $c\bar{s}$-mesons 
in four channels of $0^{-}, 1^{-}, 0^{+}, 1^{+}$, respectively, 
and similar relations for the $c\bar{n}$ mesons are also easily obtained.
Our criteria requested in the analysis of BSR are given
in sec.~IV and numerical results obtained thus are also shown. 
Finally, summary and discussion are devoted in sec.~V.

\section{Charmed-meson spectroscopy}
\begin{figure}[h]
\vspace*{0cm}
\begin{center}
\hspace*{0cm}
\psfig{file=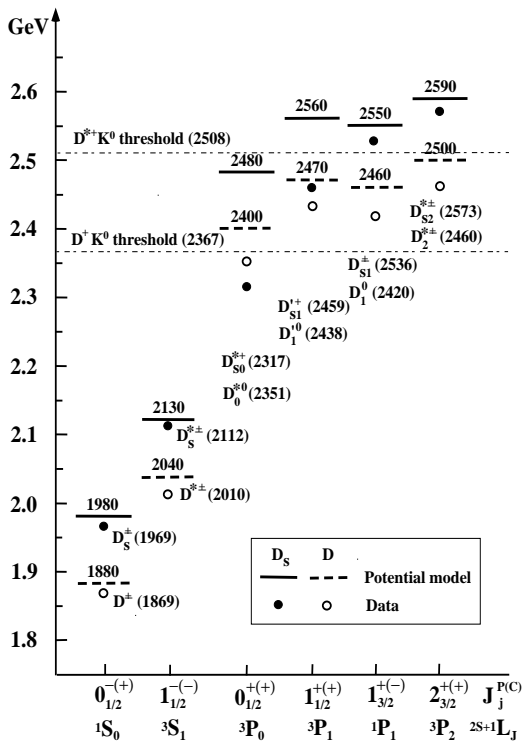,width=7cm,height=10cm}
\end{center}
\vspace{0cm}
\caption{Spectroscopy of $c\bar{n}$ and $c\bar{s}$ mesons with $L=0,1$.
The solid lines for the $c\bar{s}$ and dashed lines 
for the $c\bar{n}$ show numerical results quoted from Ref.~\cite{GK} 
and the experimental data to be compared 
are given by closed and open circles 
\cite{PDG,BABAR1,BABAR2,CLEO1,CLEO2,BELLE1,BELLE2,FOCUS1,FOCUS2}, 
respectively. Their values on the masses are written down in this figure. 
For the purpose of reference, two dash-dotted lines 
display the thresholds of $D^+K^0$ and $D^{*+}K^0$ 
decays for the $c\bar{s}$ mesons.}
\label{fig:fig1}
\end{figure}

%
In general, the classification of mesons containing 
a single heavy quark is interpreted with the help 
of heavy-quark symmetry (HQS) \cite{BS}, {\it i.e.} 
the symmetry valid for the infinitely heavy mass of charm quark. 
Under this symmetry, the strong interaction conserves 
total angular momentum of the light quark, $j$. 
In the meanwhile, total angular momentum of the light-heavy system,
$J$, should be still regarded as a good quantum number 
of the system, even if the HQS breaks down. 
Indeed, the charm quark is much heavier 
than the QCD scale ($\Lambda_{\rm QCD}\simeq 0.25$ GeV) 
but not infinitely heavy. 
In this way, we explain the classification of charmed mesons 
in terms of ($L, S, J, j$) appropriately according to circumstances. 
Here $L$ and $S$ denote the orbital angular momentum 
between the light and heavy quarks and total spin of the system, 
respectively. 
For instance, two ground states ($L=0, J^{P(C)}=0^{-(+)},1^{-(-)}$) 
form the $j=1/2$ doublets 
and four the first excited states 
($L=1, J^{P(C)}=0^{+(+)},1^{+(+)},1^{+(-)},2^{+(+)}$) 
can be understood as $j=1/2$ doublets ($J^{P(C)}=0^{+(+)},1^{+(+)}$) 
and $j=3/2$ doublets ($J^{P(C)}=1^{+(-)},2^{+(+)}$).
The states up to $L=1$ with two classifications are shown 
in the horizontal axis of Fig.~\ref{fig:fig1}.

Four of these six states, {\it i.e.} $j=1/2$ doublets 
at $L=0$ and $j=3/2$ doublets at $L=1$, have been observed 
over the past two decades ($\sim 1975-1994$) following the discovery 
of open charms \cite{BS}, because their states have relatively 
narrow widths \cite{PDG}. 
The properties of masses and widths for both charmed non-strange 
and strange mesons match well with the predictions 
from the potential model analyses \cite{GI,GK,DE}.
Recently ($2000\sim$), the leading candidates for two missing states, 
$j=1/2$ doublets at $L=1$, have been also discovered: 
the first observation of $D_1^0$ with the features of
$J^P=1^{+}$ was reported by the CLEO Collaboration \cite{CLEO2} 
and it was also observed by the BELLE Collaboration \cite{BELLE2}. 
The $D_1^0$ has the mass $M_{D_1^0}\simeq 2423$ MeV 
and the very broad width $\Gamma_{D_1^0}\simeq 329$ MeV, 
where we adopted the average of the above two data.
Besides the $D_1^0$, the BELLE Collaboration observed 
also a very broad resonance $D_0^{*0}$ 
with the features of $J^P=0^{+}$ \cite{BELLE2}. 
Though a marginally compatible result with that of the BELLE for the mass, 
the FOCUS Collaboration also confirmed the evidence of the
$D_0^{*0}$ and at the same time observed its charged partner
$D_0^{*+}$, which also has the rather broad width \cite{FOCUS2}. 
The average of the BELLE and FOCUS data on the $D_0^{*0}$ 
gives $M_{D_0^{*0}}\simeq 2351$ MeV and  $\Gamma_{D_0^{*0}}\simeq 262$
MeV. The FOCUS data on the $D_0^{*+}$ 
is $M_{D_0^{*+}}\simeq 2403$ MeV 
and $\Gamma_{D_0^{*+}}\simeq 283$ MeV.
As seen from these data, the $j=1/2$ doublets at $L=1$ have the very broad widths 
of several hundreds MeV in contrast with the $j=3/2$ doublets at $L=1$. 
Their masses and broad widths are not in conflict
with predictions from the potential model analyses \cite{GK,DE}.
Very recently, also for the charmed-strange mesons 
four experimental groups (BABAR, CLEO, BELLE, FOCUS) 
independently have observed the leading candidates 
for missing $j=1/2$ doublets at $L=1$. 
Contrary to our natural speculation about their masses 
and widths in analogy with the charmed non-strange mesons, 
surprising facts were reported by the observations of their groups, 
as mentioned below in detail.

The $D_{s0}^{*+}(2317)$ was first discovered 
in the $D_s^+\pi^0$ decay channel 
with the very narrow width by the BABAR Collaboration
\cite{BABAR1}, and in the same decay channel its existence 
was soon later confirmed by the CLEO~\cite{CLEO1} 
and the BELLE~\cite{BELLE1} Collaborations. 
Its mass was measured to be $2317$ MeV commonly, 
which is approximately $46$ MeV below the $DK$ threshold. 
The width was determined to be less than $4.6$ MeV, 
which is compatible with the experimental resolution. 
Here we adopted the severest upper bound from the BELLE \cite{BELLE1}. 
More recently, the FOCUS Collaboration \cite{FOCUS1} reported 
a preliminary result on the measurement of the mass $2323$ MeV, 
slightly larger than those from the other three experiments.
Together with the $D_{s0}^{*+}(2317)$, the CLEO \cite{CLEO1} 
and the BELLE \cite{BELLE1} observed another narrow resonance 
in the $D_s^{*+}\pi^0$ decay channel, with the mass close to $2.46$
GeV, where we denote the state by $D_{s1}^{\prime +}$. 
Later, the BABAR confirmed its existence 
in the same decay channel \cite{BABAR2}, although they have 
already found such a signal in Ref.~\cite{BABAR1}. 
It can be assigned to $J^P=1^+$ from the observation of 
the radiative decay in $D_s^+ \gamma$ and the analysis of 
helicity distributions. Their average mass is $2459$ MeV, 
which is approximately $46$ MeV below $D^{*+}K^0$ 
and the width is less than $5.5$ MeV, which is the severest 
upper bound observed by the BELLE~\cite{BELLE1}.

On the other hand, prior to such observations, 
most potential models as typified by Refs.~\cite{GK,DE} 
have predicted that two missing states of $0^+$ and $1^+$ 
($L=1,j=1/2$ doublets) should be massive enough 
to decay into $DK$ and $D^*K$ in a $s$-wave, respectively. 
The widths are then expected to be very broad due to the strong decays. 
The spectroscopy of $c\bar{n}$ and $c\bar{s}$ mesons with $L=0,1$ is drawn 
in Fig.~\ref{fig:fig1} in order to compare theoretical predictions 
(solid lines for the $c\bar{s}$ and dashed lines for the $c\bar{n}$) 
with the experimental data (closed circle for the $c\bar{s}$ 
and open circle for the $c\bar{n}$). We refer to numerical results 
in Ref.~\cite{GK} as a typical calculation by the potential models. 
Their values on the masses are written down in the figure. 
%
%
After the observations, using data of their new states, 
Cahn and Jackson \cite{CJ} showed that there are no model 
parameters to explain all the masses and widths 
of the $0^+$ and $1^+$ states consistently within the potential models.
Thus, large discrepancies between the observations 
and the theoretical predictions for the masses 
and the widths have renewed theoretical interests 
in the issue of the charmed-strange meson spectroscopy, 
especially for the $0^+$ state \cite{comment1}. 
Such a large discrepancy of the mass in the $0^+$ channel
is seen also in the QCD sum rule analysis based on HQET \cite{DHLZ} and 
lattice QCD calculations in the static limit \cite{Bali} or with dynamical quarks \cite{UKQCD}, although their analyses still have large errors. 
To resolve this issue, a lot of theoretical ideas 
have been suggested. These ideas would be categorized 
broadly into two cases: modification of prior conventional 
models or claim of exotic states different from prior conventional structures.
As one of improvements within the context of conventional 
quark models, it was shown that the $DK$ mixing 
with the $p$-wave $c\bar{s}$ state plays an important role 
to lower the mass \cite{vBR}. 
Apart from such a conventional picture, 
a possibility of the exotic configuration has been suggested: 
a four-quark state \cite{CH,Terasaki,HT}, $DK$ 
molecule \cite{BCL}, $D_s\pi$ atom \cite{Szczepaniak}, 
and four-quark states with a $DK$ mixing \cite{BPP}. 
Using perturbative QCD, Chen and Li calculated 
branching ratios of $D_{s0}^{*+}(2317)$ productions 
in the $B$-meson decays and showed that the BELLE measurement 
of $B$-meson decays favors exotic multi-quark states 
of the $D_{s0}^{*+}(2317)$, rather than conventional 
pictures \cite{CL}. However, it was argued that four-quark 
states could not be bound for the charmed-strange mesons \cite{Lipkin}.
Thus, the theoretical verdict on this crucial issue 
is still not obtained.

Another remarkable property of light-heavy systems 
is that in the heavy quark limit the $j=1/2$ doublets 
at $L=1$ could be interpreted as a chiral partner of 
those at $L=0$ \cite{NRZ,BEH}. According to this idea, 
one could expect that the mass-splittings 
between the $0^+(1^+)$ and $0^-(1^-)$ states are 
almost equivalent to $m_N/3$, where $m_N$ is a nucleon mass. 
The masses of the $0^+$ and $1^+$ states obtained from 
this value are indeed near the values measured by experiments.
This property requires us comprehensive discussion 
involving not only the excited states of $0^+$ and $1^+$ 
but also the ground states of $0^-$ and $1^-$, 
when we investigate the masses of $0^+$ and $1^+$ states in detail.

Motivated by these, we formulate the QSRs
and show the Borel-transformed forms for eight states of the $c\bar{s}$ and $c\bar{n}$ mesons with $J^{PC}=0^{-}, 1^{-}, 0^{+}, 1^{+}$ in the next section.
Note that we study these eight states independently in all the same framework and criteria for the BSR, and discuss their relative energy levels.

\section{Formulation of Borel sum rules}
We start by considering the covariant 
two-point function $\Pi_R(q^2)$ with quark currents 
$J_R(x)$ ($R=0^-,1^-,0^+,1^+$) \cite{SVZ,RRYpr}
\beq
\Pi_R(q^2)=i\int d^4x e^{iq\cdot x} \la 0|T J_R(x) J_R^\dagger(0) |0\ra,
\label{eqn:qsr0-1}
\eeq
where $|0\ra$ is nonperturbative QCD vacuum 
and $T$ the time-ordered product. 
$J_R(x)$ denotes the pseudo-scalar, vector, scalar 
and axial-vector currents described respectively as
\beq
J_{0^-}(x)&=& i\bar{q}(x)\gamma_5 c(x),
\label{eqn:qsr1-1}\\
J^{\mu}_{1^-}(x)&=& \bar{q}(x)\gamma^\mu c(x),
\label{eqn:qsr1-2}\\
J_{0^+}(x)&=& \bar{q}(x) c(x),
\label{eqn:qsr1-3}\\
J^{\mu}_{1^+}(x)&=& \eta^{\mu\nu}\bar{q}(x)\gamma_\nu\gamma_5 c(x),
\label{eqn:qsr1-4}
\eeq
with $q(x)$ being light-quark ($d$ or $u$ 
under the isospin symmetry) fields or strange-quark ($s$) field and
$c(x)$ being charm-quark field at the point $x$, and 
$\eta^{\mu\nu}=q^\mu q^\nu/q^2-g^{\mu\nu}$. Note that, {e.g.}, 
the Lorentz structure of the vector-current correlator 
with $J^{\mu}_{1^-}(x)$ and $J^{\nu\dagger}_ {1^-}(0)$ 
can be generally expressed as the sum of the transverse invariant 
function proportional to $q^\mu q^\nu/q^2-g^{\mu\nu}$ 
and the longitudinal invariant function proportional 
to $q^\mu q^\nu/q^2$. We here confine our attention 
to only the transverse component. In the axial-vector case, 
automatically only the transverse component remains 
owing to addition of Lorentz tensor $\eta^{\mu\nu}$ 
and hence we can get rid of the contribution 
from the pseudo-scalar meson in the lowest state \cite{HKL1}.

We define invariant functions $\Pi_i(q^2)$ ($i=P,V,S,A$), 
which are directly evaluated in our QSR, 
for respective channels as follows:
$\Pi_{0^-}(q^2)=\Pi_P(q^2)$, 
$\Pi^{\mu\nu}_{1^-}(q^2)=(q^\mu q^\nu/q^2-g^{\mu\nu}) \Pi_V(q^2)$, 
$\Pi_{0^+}(q^2)=\Pi_S(q^2)$ and 
$\Pi^{\mu\nu}_{1^+}(q^2)=(q^\mu q^\nu/q^2-g^{\mu\nu})\Pi_A(q^2)$.
Under these definitions, the dispersion 
relation satisfied by such invariant functions 
is given without any subtraction as
\beq
\Pi(q^2)=\frac{1}{\pi}
\int ds \frac{{\rm Im}\Pi(s)}{s-q^2-i\epsilon}.
\label{eqn:qsr0-2}
\eeq
This equation relates the real part of the correlation function, 
which is valid in the high momentum region ($-q^2\rightarrow \infty$), 
with the imaginary part, which is valid 
in the low momentum region ($q^2\rightarrow \Lambda_{QCD}^2$) 
with a typical soft-scale $\Lambda_{QCD}$.
Via this dispersion relation, we construct the QSR 
by equating the side of operator product expansion (OPE) 
and the phenomenological (PH) side related to the spectral function. 
The former is described as the product of Wilson coefficients 
and nonperturbative QCD vacuum condensates or quark masses, 
while the latter is parametrized by hadronic quantities 
such as resonance masses, couplings and the continuum 
threshold, {\it etc.} Thus, the QSR is represented in a simple form,
\beq
\int_{m_c^2}^\infty ds\ W(s) \frac{1}{\pi}
\left({\rm Im}\Pi^{PH}(s)- {\rm Im}\Pi^{OPE}(s)\right)=0,
\label{eqn:qsr0-3}
\eeq
where $W(s)$ is an arbitrary weight function, 
but must be analytic except for the positive real axis. 
The lower limit of the integration is given by the charm-quark 
Mass squared, $m_c^2$. We adopt $W(s)={\rm exp}(-s/M^2)$ in the BSR below, 
where $M$ is known as an artificial so-called Borel mass.
We take the most general ansatz for the spectral function 
of the PH side, {\it i.e.}, it would be saturated 
by one resonance in the narrow width limit 
and a continuum in the form of a step function:
\beq 
\lefteqn{\frac{1}{\pi}{\rm Im}\Pi^{PH}(s)}
\nonumber\\
&=& F \delta(s-m_R^2)+ \frac{1}{\pi}{\rm Im}\Pi^{OPE}(s)\,\theta(s-s_R)
\label{eqn:qsr0-4}
\eeq
with the QCD continuum threshold $s_R$. $F$ is a pole residue, 
where we take $F=f_R^2 m_R^{2k}$ with $f_R$ the couplings 
of the lowest resonances with respective parities to the 
hadronic current and $m_R$ a pole mass. The power $k$ 
of $m_R^2$ in the pole residue is taken to match 
the maximum power of $s$ in the asymptotic $s$-behavior 
of the spectral function. 
For $s > s_R$, we assume that the hadronic continuum reduces 
to the same form with that obtained by an analytic continuation 
of the OPE, i.e. the perturbative terms, based on a hypothesis 
of the quark-hadron duality. Such an assumption is expected 
to smear the contributions of the higher radial excitations.
This simple ansatz would indeed reproduce well the characteristic 
features of the real spectrum in most applications of sum rules 
to mesons and baryons \cite{RRYpr}.

Next, we display analytic forms of the QSR in each channel 
by implementing the Borel transformation to the QSR 
as in Eq.~(\ref{eqn:qsr0-3}) \cite{NSVZ}.
Performing the OPE at dimension $d\le 6$ operators, 
we obtain the relations of the BSR 
for four channels ($0^+, 0^-, 1^+, 1^-$) of the $c\bar{s}$ meson: 
in the scalar $0^+$ (pseudoscalar $0^-$) channels, those read
\beq
\lefteqn{f_{0^\pm}^2 m_{0^\pm}^2 e^{-m_{0^\pm}^2/M^2}}
\nonumber\\
&=&\frac{3}{8\pi^2}\int_{m_c^2}^{s_{0^\pm}}
ds\ e^{-s/M^2}s\left(1-\frac{m_c^2}{s}\right)^2 
\nonumber\\
&\times& \left(1\mp \frac{2 m_c m_s}{s-m_c^2}
+\frac{4}{3}\frac{\alpha_s(s)}{\pi}R_0(m_c^2/s)\right)
\nonumber\\
&+& e^{-m_c^2/M^2}
\left[\pm m_c\la\bar{s}s\ra_0
+\frac{1}{2}\left(1+\frac{m_c^2}{M^2}\right)m_s\la\bar{s}s\ra_0
\right.
\nonumber\\
&+& \frac{1}{12}\left(\frac{3}{2}-\frac{m_c^2}{M^2}\right)
\left\la\frac{\alpha_s}{\pi}G^2\right\ra_0
\nonumber\\
&\pm& \frac{1}{2}\frac{1}{M^2}\left(1-\frac{1}{2}\frac{m_c^2}{M^2}\right)
m_c \la\bar{s}g\sigma\cdot G s\ra_0
\nonumber\\
&-& \frac{1}{12}\frac{m_c^4}{M^6}m_s\la\bar{s}g\sigma\cdot G s\ra_0
\nonumber\\
&-& \left.\frac{16\pi}{27}\frac{1}{M^2}
\left(1+\frac{1}{2}\frac{m_c^2}{M^2}-\frac{1}{12}\frac{m_c^4}{M^4}\right)
\alpha_s\la\bar{s}s\ra_0^2\right],
\label{eqn:bsr1-1}
\eeq
where $g$ denotes the strong coupling, 
$G^2=G_{\mu\nu}G^{\mu\nu}$ with gluon field $G_{\mu\nu}$ and
$\sigma \cdot G=\sigma_{\mu\nu}G^{\mu\nu}$.
$\la O \ra_0$ with the composite operator $O$ 
represents $\la 0|O(0)|0\ra$ with the local composite 
operator $O(0)$ at the origin. 
On the other hand, in the axial-vector $1^+$ (vector $1^-$) 
channels, we similarly obtain
\beq
\lefteqn{f_{1^\pm}^2 m_{1^\pm}^2 e^{-m_{1^\pm}^2/M^2}}
\nonumber\\
&=&\frac{1}{8\pi^2}\int_{m_c^2}^{s_{1^\pm}}
ds\ e^{-s/M^2}s\left(1-\frac{m_c^2}{s}\right)^2 \left(2+\frac{m_c^2}{s}\right)
\nonumber\\
&\times& \left(1\mp \frac{3 m_c m_s s}{(2s+m_c^2)(s-m_c^2)}
+\frac{4}{3}\frac{\alpha_s(s)}{\pi}R_1(m_c^2/s)\right)
\nonumber\\
&+& e^{-m_c^2/M^2}
\left[\pm m_c\la\bar{s}s\ra_0
+\frac{1}{2}\frac{m_c^2}{M^2}m_s\la\bar{s}s\ra_0
\right.
\nonumber\\
&-& \frac{1}{12}\left\la\frac{\alpha_s}{\pi}G^2\right\ra_0
\mp \frac{1}{4}\frac{m_c^2}{M^4}
m_c \la\bar{s}g\sigma\cdot G s\ra_0
\nonumber\\
&+& \frac{1}{12}\frac{1}{M^2}\left(1+\frac{m_c^2}{M^2}-\frac{m_c^4}{M^4}\right)
m_s\la\bar{s}g\sigma\cdot G s\ra_0
\nonumber\\
&-& \left.\frac{20\pi}{81}\frac{1}{M^2}
\left(1+\frac{m_c^2}{M^2}-\frac{1}{5}\frac{m_c^4}{M^4}\right)
\alpha_s\la\bar{s}s\ra_0^2\right].
\label{eqn:bsr1-2}
\eeq
Here, we approximate at the first order of $m_s$, 
because $m_s$ is small enough compared to $m_c$ 
or a typical scale of $M$ ($M\sim 1$ GeV).
The $O(\alpha_s)$ corrections to the perturbative 
contributions are given as the function $R_0(x)$ 
for the (pseudo-) scalar channel \cite{RRY1,Narison1} 
and the function $R_1(x)$ for the (axial-) vector channel 
\cite{RYR,Narison1} as follows:
\beq
R_0(x)&=&\frac{9}{4}+2Li_2(x)+{\rm ln}x\,{\rm ln}(1-x)
-\frac{3}{2}\,{\rm ln}\frac{1-x}{x}
\nonumber\\
&-&{\rm ln}(1-x)+x\,{\rm ln}\frac{1-x}{x}
-\frac{x}{1-x}{\rm ln}x
\label{eqn:ope2-1}\\
R_1(x)&=&\frac{13}{4}+2Li_2(x)+{\rm ln}x\,{\rm ln}(1-x)
-\frac{3}{2}\,{\rm ln}\frac{1-x}{x}
\nonumber\\
&-&{\rm ln}(1-x)+x\,{\rm ln}\frac{1-x}{x}
-\frac{x}{1-x}{\rm ln}x
\nonumber\\
&+&\frac{(3+x)(1-x)}{2+x}\,{\rm ln}\frac{1-x}{x}
-\frac{2x}{(2+x)(1-x)^2}\,{\rm ln}x
\nonumber\\
&-&\frac{5}{2+x}-\frac{2x}{2+x}-\frac{2x}{(2+x)(1-x)}
\label{eqn:ope2-2}
\eeq
with the Spence function $Li_2(x)=-\int_0^x dt\,t^{-1}{\rm ln}(1-t)$.
As for the running coupling constant $\alpha_s(s)$ 
appearing in the perturbative terms of 
Eqs.~(\ref{eqn:bsr1-1}) and (\ref{eqn:bsr1-2}), 
we approximate it by a one-loop form, 
$\alpha_s(s)=4\pi/(9{\rm ln}(s/\Lambda_{QCD}^2))$ 
with $\Lambda_{QCD}^2=(0.25\,{\rm GeV})^2$, 
which is determined to reproduce $\alpha_s(1\,{\rm GeV})\simeq 0.5$ \cite{LBL}.

As easily seen from Eqs.~(\ref{eqn:bsr1-1}) 
and (\ref{eqn:bsr1-2}), we can get the relations 
of the OPE in the $0^-$ and $1^-$ channels 
by replacing simply $m_c$ by $-m_c$ for those in the $0^+$ and $1^+$.
Also, replacing $m_s$ and $\bar{s}s$ by $m_n$ 
and $\bar{n}n$ respectively, we can derive 
the corresponding BSRs for the non-strange $c\bar{n}$ mesons 
except for the only difference of the continuum thresholds. 
We here take the massless limit ($m_n=0$) of light quarks 
for the BSR of the $c\bar{n}$ mesons. 
These results indeed reproduce the corresponding BSRs 
for light mesons in the vector, scalar and axial-vector 
channels (see the results of Ref.~\cite{HKL2} in vacuum).
Looking on the relations of OPE in the right-hand sides 
of (\ref{eqn:bsr1-1}) and (\ref{eqn:bsr1-2}), 
we find that the difference of property between the $c\bar{n}$ 
and $c\bar{s}$ mesons would be caused by the effect of strange 
quarks that is mainly the difference between lowest 
dimensional condensates, $m_c\left\la \bar{n}n \right\ra_0$ 
and $m_c\left\la \bar{s}s \right\ra_0$.
For the pseudoscalar channel of the $c\bar{n}$ mesons, the similar relation 
was used in Ref.~\cite{AE}, although their relations are somewhat 
different from ours for whole sign of a term including 
gluon condensate and a part of terms including four-quark
condensates. This difference, however, is insensitive to the final results.
For the axial and vector channels of the $c\bar{n}$ mesons, 
under approximation neglecting gluon condensates 
and four-quark condensates, the same relation was used 
in Ref.~\cite{Eletsky}. More recently, Narison presented 
explicit BSR forms in the scalar and pseudoscalar channels 
of the $c\bar{n}$ and $c\bar{s}$ mesons \cite{Narison2}.
They are very similar to our forms.

In order to extract the values of physical quantities
from the BSR, we use the following standard values 
for QCD parameters appearing in the OPE:
$m_s=0.12$ GeV \cite{RRYpr}, $m_c=1.46$ GeV \cite{mc},
$\la\bar{n}n\ra_0=(-0.225\,{\rm GeV})^3$ \cite{RRYpr}, 
$\la\bar{s}s\ra_0=0.8\times\la\bar{n}n\ra_0$ GeV$^3$ \cite{RRYpr}, 
$\left\la\frac{\alpha_s}{\pi}G^2 \right\ra_0=(0.33\,{\rm GeV})^4
\times 2$ \cite{Narison1},
$\la\bar{n}g\sigma\cdot G n \ra_0=M_0^2\la\bar{n}n\ra_0$ 
with $M_0^2=0.8$ GeV$^2$ \cite{JCFG},
$\la\bar{s}g\sigma\cdot G s \ra_0=M_0^2\la\bar{s}s\ra_0$,
$\alpha_s\left\la\bar{n}n\right\ra_0^2$
$=0.162\times 10^{-3}$ GeV$^6$ and 
$\alpha_s\left\la\bar{s}s\right\ra_0^2$
$=0.8^2 \times \alpha_s\left\la\bar{n}n\right\ra_0^2$ GeV$^6$.
Note that we evaluated four-quark condensates 
in terms of $\left\la\bar{n}n\right\ra_0$,
supposing that the factorization or vacuum saturation 
in an intermediate state of four-quark matrix element 
can work well in a good approximation \cite{SVZ}. 
Although the ambiguity of the absolute value of the 
four-quark condensate still remains large as discussed 
by several authors (see Refs.~\cite{Narison1,JCFG}), 
the detailed discussion is not much relevant 
to our present analysis. It is because the contribution 
from higher dimensional condensates is negligible 
compared to the lowest dimensional operators ($d=3$) 
and thus the final results are independent of details 
of four-quark condensates.

Finally, taking a logarithmic derivative with respect 
to $1/M^2$ for both sides of (\ref{eqn:bsr1-1}) 
and (\ref{eqn:bsr1-2}), we get rid of $f_R^2$-dependence 
completely from their equations and thus can derive 
the Borel curve of the resonance masses $m_R^2$ as 
functions of $M^2$ and $s_R$.

\section{Numerical results}
We determine plausible values of the resonance mass extracted 
from the Borel curves in accordance with the following criteria. 
Its method is in order:
\begin{enumerate}
\item We plot the Borel curves of the resonance 
mass $m_R$ as a function of $M$ with taking 
some selected values of the continuum thresholds $\sqrt{s_R}$. 
In general, the values of the thresholds might be different 
from the phenomenological value of the first radial 
excitation mass and rather would be smaller. This is natural, 
because our simple ansatz in the PH side 
of the QSR cannot take into account the complicated 
structure of the resonances in this region. We, therefore, 
set the maximum value of $\sqrt{s_R}$, $\sqrt{s_R^{max}}$, to be the mass
of the first radial excitation predicted by the potential model \cite{DE}. 
\item In the OPE side, the contribution of the nonperturbative 
corrections should be generally small compared to that 
of the perturbative terms in large momentum region. 
Since we, however, truncate the OPE at $d=6$ by hand, 
we would make the impact of this approximation as small as possible. 
In naive sense it is satisfied requiring that 
the contribution from higher dimensional operators than $d=6$ 
becomes thoroughly smaller than the perturbative terms 
in large region of $M$. In such a region of $M$, the convergence 
of the OPE can indeed be ensured to some extent. 
Here, we quantitatively require a constraint that 
the effect of $d=6$ term is less than $20\%$ compared 
to that of the leading terms in Eqs.~(\ref{eqn:bsr1-1}) 
and (\ref{eqn:bsr1-2}). This assures us 
to get physical quantities with about $10\sim 20\%$ accuracy. 
This condition gives the lower limit of $M$, $M_{min}$.
\item In the PH side, 
we would make the continuum contribution as small as possible,
since we are interested only in the pole mass.
It would be possible by taking small region of $M$. 
We require that the continuum contribution is less than $20\%$ 
compared to that of the perturbative contribution. 
For the leading terms in Eqs.~(\ref{eqn:bsr1-1}) 
and (\ref{eqn:bsr1-2}), the continuum contribution 
corresponds to the integration over $s$ from $s_R$ 
to infinity, while the perturbative contribution 
has the integration from $m_c^2$ to infinity. 
This constraint also assures us to get physical quantities 
with about $10\sim 20\%$ accuracy. From this condition, 
we can give the upper limit of $M$, $M_{max}$, 
which strongly depends on the threshold in contrast to $M_{min}$.
\item Thus, we set the Borel window, $M_{min}< M < M_{max}$, 
so that the BSR would work well within the window  
and we extract more reliable values of physical quantities 
from the Borel curves.
\item For $\sqrt{s_R}<\sqrt{s_R^{max}}$, 
we look for the stable regions (Borel stability) of the Borel curves 
vs. $M$ within the Borel windows. 
Then, from the curve having the widest Borel stability, 
we read off the resonance mass 
and the threshold as the most reliable values of physical quantities. 
In general, this would give the central values 
of physical quantities under consideration. 
Finally, we adopt all resonance masses 
and thresholds to the extent that the Borel stabilities 
on the curves exist within the Borel windows. 
This would give errors for the central values. 
Note that in actual evaluation for the Borel stability 
we do not mind its accurate positions 
and widths of the stable regions. 
Such an ambiguity also should be included in the above errors.
\end{enumerate}

\begin{figure}[h]
\begin{center}
\psfig{file=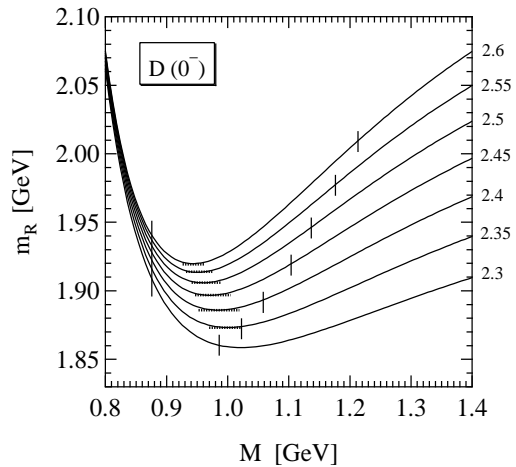,width=7.0cm,height=6.5cm}
\end{center}
\vspace{-0.5cm}
\caption{The Borel curves of $m_{D(0^-)}$ vs.
$M$, where $\sqrt{s_{D(0^-)}}=2.3\sim 2.6$ GeV 
at $0.05$ GeV intervals. Explanation for the lines
is given in the text.}
\label{fig:D_pseudo}
\end{figure}

Following the above criteria, 
we plot the Borel curves of resonance masses $m_R$ 
vs. $M$ in two ground states ($R=0^-,1^-$) of the $c\bar{n}$ mesons.
We first study these two channels 
to determine a size of charm-quark mass $m_c$, 
which is treated as an input data in the OPE side. 
Since the quark mass is a scale-dependent quantity, 
its size runs with the energy scale of the system, {\it e.g.} 
we adopt $m_c=1.2 \sim 1.5$ GeV in the QSR approach 
\cite{RRY2,Narison3}. We determine $m_c$ to reproduce 
the masses of both states as well as possible. 
This leads to $m_c=1.46$ GeV as listed before.
%
\begin{figure}[h]
\begin{center}
\psfig{file=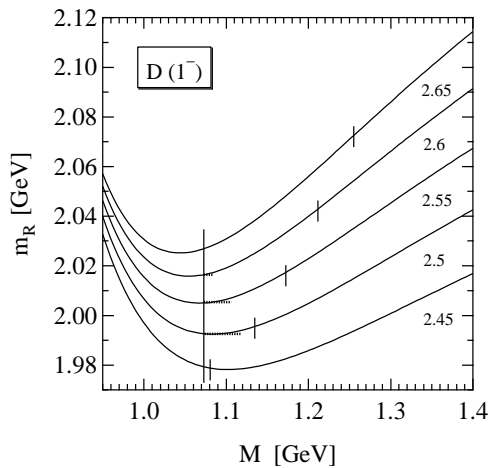,width=7.0cm,height=6.5cm}
\end{center}
\vspace{-0.5cm}
\caption{The Borel curves of $m_{D(1^-)}$ vs. $M$, where 
$\sqrt{s_{D(1^-)}}=2.45\sim 2.65$ GeV at $0.05$ GeV intervals.}
\label{fig:D_vector}
\end{figure}
%
Using this input, we shows the curves 
for the $0^-$ channel with $\sqrt{s_{D(0^-)}}=2.3 \sim 2.6$ GeV 
at $0.05$ GeV intervals in Fig.~\ref{fig:D_pseudo}  
and for the $1^-$ channel 
with $\sqrt{s_{D(1^-)}}=2.45 \sim 2.65$ GeV at $0.05$ GeV intervals
in Fig.~ \ref{fig:D_vector}.
%
\begin{figure}[h]
\begin{center}
\psfig{file=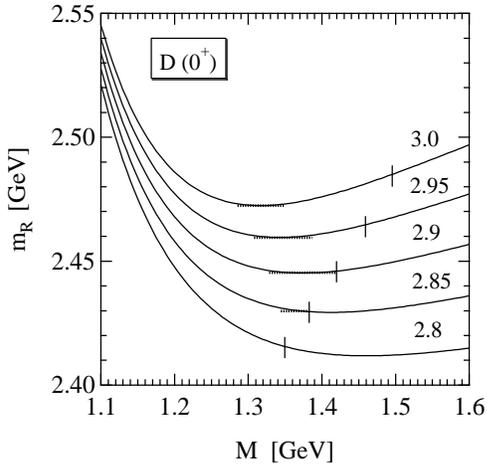,width=7.0cm,height=6.5cm}
\end{center}
\vspace{-0.5cm}
\caption{The Borel curves of $m_{D(0^+)}$ vs. $M$, 
where $\sqrt{s_{D(0^+)}}=2.8\sim 3.0$ GeV 
at $0.05$ GeV intervals. The lower limits of the Borel windows 
are below $M=1.1$ GeV.}
\label{fig:D_scalar}
\end{figure}
%
In Fig.~ \ref{fig:D_pseudo}, left long (right short) 
vertical-lines on the curves represents the lower (upper) limit 
of the Borel window, $M_{min}(M_{max})$, 
which very weakly decreases (strongly increases) with the threshold.
We show the Borel stabilities by dotted lines 
within these Borel windows. The length of 
the lines corresponds to the range of stable regions.
%
\begin{figure}[h]
\begin{center}
\psfig{file=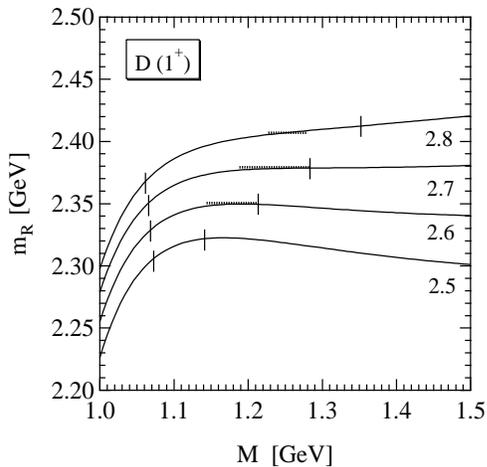,width=7.0cm,height=6.5cm}
\end{center}
\vspace{-0.5cm}
\caption{The Borel curves of $m_{D(1^+)}$ vs. $M$, 
where $\sqrt{s_{D(1^+)}}=2.5\sim 2.8$ GeV at $0.1$ GeV intervals.}
\label{fig:D_axial}
\end{figure}
%
In the $0^-$ channel, we cannot find 
any Borel stability below $\sqrt{s_{D(0^-)}}=2.3$ GeV, 
while above the value there exist such stable regions 
on the curves until the upper limit $\sqrt{s_{D(0^-)}}=2.6$ GeV. 
Thus, we get the resonance mass, $m_{D(0^-)}=1.90\pm 0.03$ GeV.
In the $1^-$ channel, we cannot find any Borel stability 
below $\sqrt{s_{D(1^-)}}=2.45$ GeV, while above the value there exist such stable regions on the curves 
until $\sqrt{s_{D(1^-)}}=2.6$ GeV, 
which is below the upper limit $\sqrt{s_{D(1^-)}}=2.7$ GeV. 
Thus, we get the resonance mass, $m_{D(1^-)}=2.00\pm 0.02$ GeV.
Indeed, the obtained results for resonance masses 
are near experimental values, $m_{D(0^-)}^{exp.}\simeq 1.867$ GeV 
and $m_{D(1^-)}^{exp.}\simeq 2.010$ GeV, respectively. 
In a similar analysis, we get the results of other channels, 
$m_{D(0^+)}=2.45\pm 0.03$ GeV and $m_{D(1^+)}=2.38\pm 0.05$ GeV, 
which are read off Figs.~\ref{fig:D_scalar}, \ref{fig:D_axial}, respectively. 
Here, we took $\sqrt{s_{D(0^+)}}=2.8\sim 3.0$ GeV and $\sqrt{s_{D(1^+)}}=2.5\sim 2.8$ GeV, respectively. 
%
\begin{table}
\begin{ruledtabular}
\caption{\label{tbl:table1}
Numerical results of the resonance mass $m_R$ 
and the continuum threshold $\sqrt{s_R}$.
They are listed for four channels ($R=0^-,1^-,0^+,1^+$) 
of the $c\bar{n}$ mesons.
We refer to Ref.~\cite{PDG,BELLE2} for the experimental data 
and the predictions from Ref.~\cite{DE} for maximum values of the thresholds.}
\vspace{0.2cm}
\begin{tabular}{ccccc}
&  & $D$ [GeV] &  &  \\
$R$   & $m_R$ & $m_R$(exp.) & $\sqrt{s_0}$ &  pt. model \cite{DE} \\ \hline
$0^-$  & 1.90 $\pm$ 0.03 & 1.869 & 2.45 $\pm$ 0.15 & 2.589 \\
$1^-$  & 2.00 $\pm$ 0.02 & 2.010 & 2.55 $\pm$ 0.05 & 2.692 \\
$0^+$  & 2.45 $\pm$ 0.03 & 2.351 & 2.90 $\pm$ 0.10 & 2.949 \\
$1^+$  & 2.38 $\pm$ 0.05 & 2.438 & 2.70 $\pm$ 0.10 & 3.045 \\
\end{tabular}
\end{ruledtabular}
\end{table}
%
We summarize these results in Table~\ref{tbl:table1}.
From the table, we find that the excited ($p$-wave) 
$1^+$ state reproduce the experimental data within errors, 
while the excited $0^+$ state seems to be overestimated 
by about 100 MeV in comparison with the experimental data. 
However, if we take into account 
the broad width ($\Gamma_{D(0^+)}\simeq 329$ MeV) measured 
by experiments, our result could be not in conflict with such data 
(within the large width). 
Also, the obtained thresholds are different among the four 
channels and thus have a level structure similar to the obtained 
resonance masses; the order of this energy level is the same 
as that of the resonance masses.

\begin{figure}[h]
\begin{center}
\psfig{file=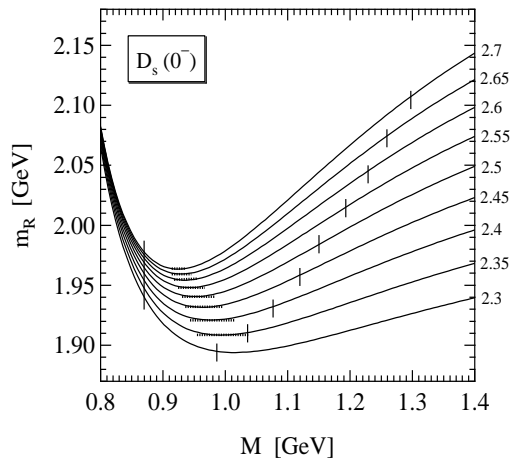,width=7.0cm,height=6.5cm}
\end{center}
\vspace{-0.5cm}
\caption{The Borel curves of $m_{D_s(0^-)}$ vs. $M$,
where $\sqrt{s_{D_s(0^-)}}=2.3\sim 2.7$ GeV at $0.05$ GeV intervals.}
\label{fig:Ds_pseudo}
\end{figure}
%
Next, we discuss the case of the $c\bar{s}$ mesons following 
the above criteria. Here we fix $m_s=0.12$ GeV 
as listed before. Indeed, final results of 
the resonance masses are almost insensitive 
to the adopted values of $m_s$. For example, the result 
with $m_s=0.15$ GeV changes by less than 0.5 $\%$ 
of that with $m_s=0.12$ GeV. Hence, the error coming 
from such a mass ambiguity is smaller than the error 
originated from the criteria mentioned above.
%
%
\begin{figure}[h]
\begin{center}
\psfig{file=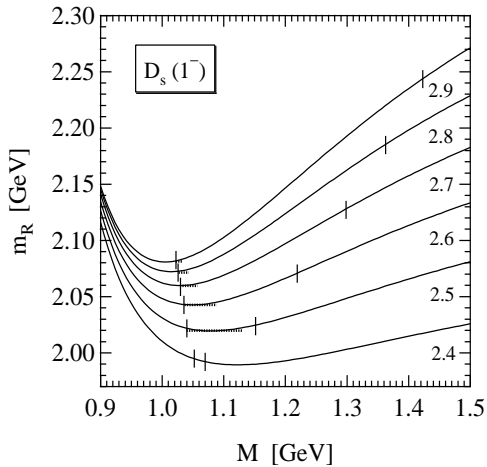,width=7.0cm,height=6.5cm}
\end{center}
\vspace{-0.5cm}
\caption{The Borel curves of $m_{D_s(1^-)}$ vs. $M$,
where $\sqrt{s_{D_s(1^-)}}=2.4\sim 2.9$ GeV at $0.1$ GeV intervals.}
\label{fig:Ds_vector}
\end{figure}
%
We show the results of the $0^-$ channel 
in Fig.~\ref{fig:Ds_pseudo}. The behaviors of the curves 
and the Borel windows are very close to those of the $D(0^-)$
in Fig.~ \ref{fig:D_pseudo}. 
We get $m_{D_s(0^-)}=1.94\pm 0.03$ GeV for $\sqrt{s_{D_s(0^-)}}=2.3\sim 2.7$ GeV 
by applying the above criteria. This value is almost consistent 
with the experimental data (1.969 GeV) within errors 
and higher by $40$ MeV than our result for the $c\bar{n}$ meson. 
Thus, our mass-splitting between the $D_s(0^-)$ and $D(0^-)$
is underestimated by about 100 MeV in comparison with 
the measured one, which would just correspond 
to a deviation between light-quark and strange-quark masses.
%
\begin{figure}[h]
\begin{center}
\psfig{file=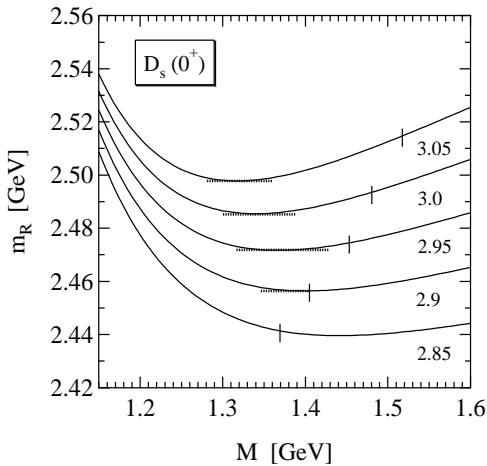,width=7.0cm,height=6.5cm}
\end{center}
\vspace{-0.5cm}
\caption{The Borel curves of $m_{D_s(0^+)}$ vs. $M$, where 
$\sqrt{s_{D_s(0^+)}}=2.85\sim 3.05$ GeV at $0.05$ GeV intervals. 
The lower limits of the Borel windows are below $M=1.15$ GeV.}
\label{fig:Ds_scalar}
\end{figure}
%
Another remarkable point is that there exists 
also a deviation between their thresholds as well 
as the mass-splitting of the resonance masses. 
This deviation is near the value of the mass-splitting. It is, 
therefore, expected that such a deviation for 
the thresholds plays an essential role to give 
the mass-splitting of the resonance masses.
%
\begin{figure}[h]
\begin{center}
\psfig{file=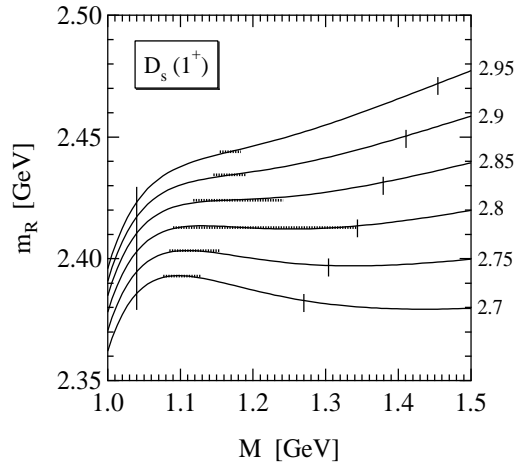,width=7.0cm,height=6.5cm}
\end{center}
\vspace{-0.5cm}
\caption{The Borel curves of $m_{D_s(1^+)}$ vs. $M$, where 
$\sqrt{s_{D_s(1^+)}}=2.7\sim 2.95$ GeV at $0.05$ GeV intervals.}
\label{fig:Ds_axial}
\end{figure}
%
We also show numerical results of other three channels 
in Figs.~
\ref{fig:Ds_vector},~\ref{fig:Ds_scalar},~\ref{fig:Ds_axial} in the same analysis. 
Every graph has the curves similar 
to the corresponding channel of the $c\bar{n}$ meson, respectively. 
In the $1^-$ channel, the obtained mass $m_{D_s(1^-)}=2.05\pm 0.04$ GeV for $\sqrt{s_{D_s(1^-)}}=2.4\sim 2.9$ GeV
is somewhat smaller than the experimental data (2.112 GeV) 
like the case of the $c\bar{n}$ meson. In this channel also, 
our mass-splitting between the $D_s(1^-)$ and $D(1^-)$ is about 
a half of the experimental observation, {\it i.e.} $\sim 50$ MeV, 
and also a difference of the obtained thresholds 
between their mesons is non-zero, {\it i.e.} about 100 MeV.
In the $0^+$ channel, we find that the obtained mass 
$m_{D_s(0^+)}=2.48\pm 0.03$ GeV for $\sqrt{s_{D_s(0^+)}}=2.85\sim 3.05$ GeV is overestimated by about 160 MeV
compared with the experimental data (2.317 GeV), 
as seen in the case of the $D(0^+)$.
This result is very close to that from the potential models \cite{GK,DE}. 
Contrary to the case of the $D(0^+)$, the $D_s(0^+)$ meson 
has been observed as the very narrow state 
($\Gamma_{D_s(0^+)}\leq 4.6$ MeV). 
Therefore, this deviation (160 MeV) is significant and 
our result supports conclusions 
from Ref.~\cite{GK,DE,CJ,Bali,DHLZ}.
A deviation between the thresholds of the $D_s(0^+)$ and $D(0^+)$ is about 100 MeV. 
In contradiction to the result in Ref.~\cite{GK,DE}, 
however, our result in the $1^+$ channel, $m_{D_s(1^+)}=2.41\pm 0.05$
GeV for $\sqrt{s_{D_s(1^+)}}=2.7\sim 2.95$ GeV, is in agreement with 
the experimental data (2.459 GeV) within errors. 
Such agreement with the data could be due to disregard of mixing 
effects between $J^{P(C)}=1^{+(+)}$ and $1^{+(-)}$, 
although Refs.~\cite{GK,DE} have taken into account the effect.
Here, our result for the threshold gives almost the same value with the case of $D(1^+)$.
These final results 
for four channels of the $c\bar{s}$ mesons 
are summarized in Table~\ref{tbl:table2}.
\begin{table}
\begin{ruledtabular}
\caption{\label{tbl:table2} 
Numerical results of the resonance mass $m_R$ 
and the continuum threshold $\sqrt{s_R}$.
They are listed for four channels ($R=0^-,1^-,0^+,1^+$) 
of the $c\bar{s}$ mesons.
For comparison, we attach experimental average values 
observed recently \cite{BABAR1,BABAR2,CLEO1,BELLE1,FOCUS1} 
in each channel and the masses of the first radial excitations,
which is predicted in Ref.~\cite{DE}.}
\vspace{0.2cm}
\begin{tabular}{ccccc}
&  & $D_s$ [GeV] &  &  \\
$R$   & $m_R$ & $m_R$(exp.) & $\sqrt{s_0}$ &  pt. model \cite{DE} \\ \hline
$0^-$  & 1.94 $\pm$ 0.03 & 1.969 & 2.50 $\pm$ 0.20 & 2.700 \\
$1^-$  & 2.05 $\pm$ 0.04 & 2.112 & 2.65 $\pm$ 0.15 & 2.806 \\
$0^+$  & 2.48 $\pm$ 0.03 & 2.317 & 3.00 $\pm$ 0.15 & 3.067 \\
$1^+$  & 2.41 $\pm$ 0.05 & 2.459 & 2.70 $\pm$ 0.25 & 3.165 \\
\end{tabular}
\end{ruledtabular}
\end{table}
%

\section{Summary and discussion}
\begin{figure}[h]
\vspace*{0cm}
\begin{center}
\hspace*{0cm}
\psfig{file=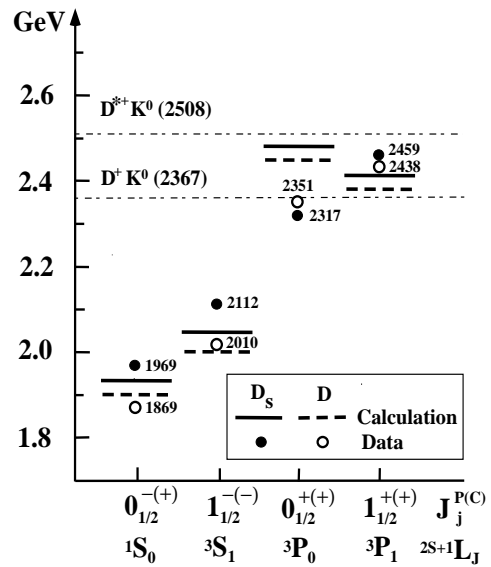,width=6.5cm,height=7.5cm}
\end{center}
\vspace{0cm}
\caption{Spectroscopy of the $c\bar{n}$ and $c\bar{s}$ mesons 
with $J^P=0^-,1^-,0^+,1^+$. The solid lines for 
the $c\bar{s}$ mesons and dashed lines for the $c\bar{n}$ mesons 
show our numerical results. The corresponding experimental 
data are given by closed and open circles, respectively, 
as in Fig.~\ref{fig:fig1}.}
\label{fig:fig2}
\end{figure}

%
We have presented detailed calculations 
of the masses for the $0^-,1^-,0^+,1^+$ charmed-mesons 
($D$, $D_s$), motivated by recent discovery 
of very narrow two states ($D_{s0}^{*+}(2317)$, 
$D_{s1}^{\prime +}(2459)$) 
by the BABAR, CLEO, BELLE and FOCUS Collaborations. 
Applying the original QSR techniques of SVZ \cite{SVZ} 
for the conventional light-heavy systems, 
we have performed the OPE up to $d=6$, involving corrections 
to the order $\alpha_s$ and to the order $m_s$ (or $m_n$). 
In this method we have taken into account full order 
of the $1/m_c$-expansion, because $m_c$ is not a large mass 
enough to approximate by the first few terms of the expansion.
We have analyzed their masses using the Borel-transformed QSR, 
because this analysis can carefully deal with the continuum contribution 
or the threshold dependence, if the resonance mass, 
which we would derive from this analysis, is sensitive 
to such a contribution \cite{comment2}. Especially, 
such care is important for discussion of the spectroscopy 
associated with (hyper-)fine structures as in our considering four channels.
In fact, as seen in Tables~\ref{tbl:table1} and \ref{tbl:table2},
we found that the mass-splittings among four channels 
of the $c\bar{n}$ or $c\bar{s}$ mesons and those in the same channels 
between the $c\bar{n}$ and $c\bar{s}$ mesons can be largely affected 
by the values of their thresholds. 
In order to derive a reliable resonance mass 
from the threshold dependence, we performed numerical calculations 
by imposing stringent criteria listed in sec.~IV for the BSR.
We compare our final results for resonance masses 
of $c\bar{n}$ and $c\bar{s}$ mesons with those of the experimental data 
in Fig.~\ref{fig:fig2}. 
%
From this figure, we find that the mass of the $D_s(0^+)$ 
meson is overestimated by about 160 MeV in comparison with 
the experimental data, which is consistent with the 
potential model analysis \cite{GK,DE}. 
Although a similar tendency is also seen 
in the mass of the $c\bar{n}$ meson in the 
same channel, these results could be consistent 
with the experimental data within the measured broad width. 
On the other hand, such a tendency is not seen
in other channels.  Conversely in other channels 
the $c\bar{s}$ masses are underestimated somewhat 
in comparison with data, independent of strange-quark mass 
adopted in our calculations. 
Hence, the $1^+ - 1^-$ and the $0^+ - 0^-$ mass-splittings 
are unequal in disagreement with the chiral multiplet 
structure predicted by chiral effective theory for heavy mesons. 
It might be expected that 
the measured low mass of $D_s^{*+}(2317)$ 
is a manifestation of any exotic states with the structure 
of a four-quarks or a molecule. 
In fact, we have seen very recently~\cite{HT} that
the assignment of the $D_{s0}^{*+}(2317)$ to the charmed-strange
four-quark meson, $\hat{F}_1$, with $(I,I_3)=(1,0)$ is favored
by the ratio of the measured rates for the 
$D_{s0}^{*+}(2317)\rightarrow D_s^{*+}\gamma$ 
to the $D_{s0}^{*+}(2317)\rightarrow D_s^+\pi^0$
decay \cite{CLEO1},
$\Gamma(D_{s}^{*+}\gamma)/\Gamma(D_s^{+}\pi^0)< 0.052$.
It is, however, hard to reconcile its assignment to an isosinglet
state (the conventional $\{c\bar{s}\}$ or an isosinglet four-quark
or a molecule) with the experimental constraint.  
Our conclusion for the charmed mesons 
with $0^+$ is different from that of Ref.~\cite{Narison2}, 
where the similar QSR analysis was performed only for $0^-$ 
and $0^+$ states of the $c\bar{s}$ and $c\bar{n}$-mesons 
without using stringent criteria like in our analysis.

In this work, we neglected renormalization-group improvement 
for the currents and the operators in the condensates, which could 
be important to improve the $Q^2$ range of the validity 
of the derived QSR. Also, as is well known, 
the physical quantities extracted from the QSR analysis 
generally have theoretical ambiguity of 
about $10\sim 20$ $\%$, which comes from the condensates 
and our criteria, {\it etc.} However, these quantitative 
ambiguities would not qualitatively change our conclusion 
as mentioned above, because we have evaluated totally eight 
states of the $c\bar{n}$ and $c\bar{n}$ mesons 
within the same method and criteria. 
In such an analysis we found an anomalous feature 
for only the charmed-strange scalar meson.

\acknowledgements
We are grateful to S.~Sasaki for fruitful discussion 
and M.~Harada for a useful comment. 
The research of A.H. is supported by the research fellowship 
from Department of Physics, Kyoto University. This work of K.T. 
is supported in part by the Grant-in-Aid for Science Research, 
Ministry of Education, Science and Culture, Japan (No. 13135101 
and No. 16540243). 




\begin{references}
\bibliographystyle{unsrt}
\setlength{\itemsep}{0.0in}
\bibitem{GI} S.~Godfrey and N.~Isgur, Phys. Rev. {\bf D32}, 189 (1985).

\bibitem{BABAR1} BABAR Collaboration, B.~Aubert {\it et al.}, 
Phys. Rev. Lett. {\bf 90}, 242001 (2003).

\bibitem{BABAR2} BABAR Collaboration, B.~Aubert {\it et al.}, 
Phys. Rev. {\bf D69}, 031101 (2004).

\bibitem{CLEO1} CLEO Collaboration, D.~Besson {\it et al.}, 
Phys. Rev. {\bf D68}, 032002 (2003).

\bibitem{BELLE1} BELLE Collaboration, K.~Abe {\it et al.}, 
Phys. Rev. {\bf D69}, 112002 (2004).

\bibitem{FOCUS1} FOCUS Collaboration, E.W.~Vaandering, hep-ex/0406044.

\bibitem{BS} J.~Bartelt and S.~Shukla, 
Ann. Rev. Nucl. Part. Sci. {\bf 45}, 133 (1995) and references therein.

\bibitem{GK} S.~Godfrey and R.~Kokoski, Phys. Rev. {\bf D43}, 1679 (1991).

\bibitem{DE} M.~Di Pierro and E.~Eichten, Phys. Rev. {\bf D64}, 114004 (2001).

\bibitem{CJ} R.N.~Cahn and J.D.~Jackson, Phys. Rev. {\bf D68}, 037502 (2003).

\bibitem{DHLZ} Y.-B.~Dai, C.-S.~Huang, C.~Liu and S.-L.~Zhu, 
Phys. Rev. {\bf D68}, 114011 (2003).

\bibitem{Bali} G.S.~Bali, Phys. Rev. {\bf D68}, 071501(R) (2003).

\bibitem{UKQCD} A.~Dougall, R.D.~Kenway, C.M.~Maynard and C.~McNeile,
the UKQCD Collaboration, Phys. Lett. {\bf B569}, 41 (2003)

\bibitem{NRZ} M.A.~Nowak, M.~Rho and I.~Zahed, 
Phys. Rev. {\bf D48}, 4370 (1993); W.A.~Bardeen and C.T.~Hill, 
Phys. Rev. {\bf D49}, 409 (1994).

\bibitem{BEH} W.A.~Bardeen, E.J.~Eichten and C.T.~Hill, 
Phys. Rev. {\bf D68}, 054024 (2003).

\bibitem{vBR} E.~van Beveren and G.~Rupp, 
Phys. Rev. Lett. {\bf 91}, 012003 (2003).

\bibitem{BCL} T.~Barnes, F.E.~Close and H.J.~Lipkin, 
Phys. Rev. {\bf D68}, 054006 (2003).

\bibitem{Szczepaniak} A.P.~ Szczepaniak, Phys. Lett. {\bf B567}, 23 (2003).

\bibitem{CH} H.-Y.~Cheng and W.-S.~Hou, Phys. Lett. {\bf B566}, 193 (2003).

\bibitem{Terasaki} K.~Terasaki, 
Phys. Rev. {\bf D68}, 011501(R) (2003); 
hep-ph/0309119; hep-ph/0309279;  hep-ph/0311069; hep-ph/0405146.

\bibitem{BPP} T.~Browder, S.~Pakvasa and A.A.~Petrov, 
Phys. Lett. {\bf B578}, 365 (2004).

\bibitem{SVZ} M.A.~Shifman, A.I.~Vainstein and V.I.~Zakharov, 
Nucl. Phys. {\bf B147}, 385, 448 (1979).

\bibitem{Narison2} S.~Narison, hep-ph/0307248.

\bibitem{PDG} Particle Data Group Collaboration, 
S.~Eidelman {\it et al.}, Phys. Lett. {\bf B592}, 1 (2004).

\bibitem{CLEO2} CLEO Collaboration, S.~Anderson {\it et al.}, 
Nucl. Phys. {\bf A663}, 647 (2000).

\bibitem{BELLE2} BELLE Collaboration, P.~Krokovny {\it et al.}, 
Phys. Rev. Lett. {\bf 91}, 262002 (2003).

\bibitem{FOCUS2} FOCUS Collaboration, J.M.~Link {\it et al.}, 
Phys. Lett. {\bf B586}, 11 (2004).

\bibitem{comment1} Indeed, the study for the $1^+$ state 
would be more complicated than that for the $0^+$ state. 
It is, {\it e.g.}, because two states such as $J^{P(C)}=1^{+(+)}, 1^{+(-)}$ 
can take place a mixing between them as mentioned in the text above.

\bibitem{HT} A.~Hayashigaki and K.~Terasaki, hep-ph/0410393.

\bibitem{CL} C.-H.~Chen and H.-n.~Li, Phys. Rev. {\bf D69}, 054002 (2004).

\bibitem{Lipkin} H.J.~Lipkin, Phys. Lett. {\bf B580}, 50 (2004).

\bibitem{RRYpr} L.J.~Rindeers, H.R.~Rubinstein and S.~Yazaki, 
Phys. Rep. {\bf 127}, 1 (1985).

\bibitem{HKL1} T.~Hatsuda, Y.~Koike and S.H.~Lee, Nucl. Phys. {\bf B394}, 221 (1993).

\bibitem{NSVZ} V.A.~Novikov, M.A.~Shifman, A.I.~Vainstein 
and V.I.~Zakharov, Fortschr. Phys. {\bf 32},11, 585 (1984).

\bibitem{RRY1} L.J.~Reinders, H.R.~Rubinstein and S.~Yazaki, 
Phys. Lett. {\bf 97B}, 257 (1980).

\bibitem{Narison1} S.~Narison, "QCD as a Theory of Hadrons, 
From Partons to Confinement", Cambridge University Press (2004) 
and references therein.

\bibitem{RYR} L.J.~Reinders, S.~Yazaki and H.R.~Rubinstein, 
Phys. Lett. {\bf 103B}, 63 (1981).

\bibitem{LBL} The value of $\alpha_s$ at any scale can be obtained from
http://www-theory.lbl.gov/~ianh/alpha/alpha.html.

\bibitem{HKL2} T.~Hatsuda, Y.~Koike and S.H.~Lee, 
Phys. Rev. {\bf C47}, 1225 (1993); 
The final analytic forms of the BSR in this paper have two trivial typos: 
first, the sign of the term with dimension-4 operators 
in the axial-vector channel should be minus and second, 
the factor of $\alpha_s$ correction in the scalar channel should be $9/3$.

\bibitem{AE} T.M.~Aliev and V.L.~Eletsky, 
Sov. J. Nucl. Phys. {\bf 38}(6), 936 (1984). 

\bibitem{Eletsky} V.L.~Eletsky, Phys. Atom. Nucl. {\bf 59}, 2002 (1996).

\bibitem{mc} This value is very close 
to a pole mass used in Ref.~\cite{Narison2,Narison3}.

\bibitem{JCFG} X.~Jin, T.D.~Cohen, R.J.~Furnstahl 
and D.K.~Griegel, Phys. Rev. {\bf C47}, 2882 (1993).

\bibitem{RRY2} L.J.~Reinders, H.R.~Rubinstein and S.~Yazaki, 
Nucl. Phys. {\bf B186}, 109 (1981).

\bibitem{Narison3} S.~Narison, Phys. Lett. {\bf B520}, 115 (2001)

\bibitem{comment2}
We also checked the resonance masses by using finite energy sum rule. 
However, this method shows strong dependence of their masses 
on the continuum thresholds and we cannot get reliable masses 
with enough accuracy like the BSR.

\end{references}
\end{document}